\documentstyle[secnumtab]{fbssuppl}

\title{An Analysis of the Use of the Cluster Separability
in Scattering Theory}
\author{M.  Damjanovi{\' c}\instnr{1}
\thanks{{\it E-mail address:} edamjano@ubbg.etf.bg.ac.yu},
Z.  Mari{\' c}\instnr{2}
\thanks{{\it E-mail address:} emaricz@ubbg.etf.bg.ac.yu}}
\instlist
{Faculty of Pharmacy, Dept. of Physics, Belgrade University,
P.O.B. 146, Vojvode Stepe 450, Yu-11000 Belgrade, Yugoslavia
\and
Institute of Physics, P.O.B. 57, Pregrevica 118, Yu-11001 Belgrade,
Yugoslavia}

\begin{document}

\maketitle

\vspace*{-10mm}
\begin{flushright}{nucl-th/9511014}\end{flushright}
\vspace*{1mm}

\begin{abstract}
In the framework of the Time Dependent Scattering Theory we discuss three
forms of Cluster Separability as well as the conditions for the
representation of the scattering system dynamics implied by their
respective use.
\end{abstract}

For a broad class of theories applied to the studies of the scattering
processes of particles and their bound states,
and in particular for the Quantum Field Theory (QFT),
one needs to include the procedure which regulates a segmentation and
clusterisation of the scattering process.
Implications of the implementation of such a procedure as well as its
relationship to the general formalism deserve a careful analysis.
In this paper we would like to point out the
relationship existing between Cluster Separability (CS), Adiabatic Cluster
Separability (ACS) and Disturbative Adiabatic Cluster Separability (DACS).
The last one being introduced by us in the previous publication
%\cite{olymp}
[1].
These three competing procedures are stated in the form of a principle for
scattering theory as conditions for the representation of
the scattering system dynamics.

The common feature of all of them is that they imply distinction between
central interaction channel of the system and two channels corresponding
to the initial and final subsystems respectively. Channels could be
described by corresponding Hamiltonians and states in their corresponding
Hilbert spaces.

What makes a difference between CS, ACS and DACS is the degree and a way of
describing the "phase
transitions" between the corresponding segments of the scattering process.

We will consider the scattering process $\alpha \rightarrow \beta$
in the framework of Time dependent propagator theory.
Initial and final channels of the process are defined with the sets of the
process constituents - particles and their bound states:
$\{ I_1, ... , I_k\} = \alpha$; $\{ F_1, ... , F_l \} = \beta$.
A corresponding Hilbert space state is denoted by $\mid\alpha>$.
The Bethe-Salpeter amplitudes
$\chi_\alpha$, ${\bar\chi}_\alpha$ for channel $\alpha$
and process propagator $G_{\alpha\beta}$ are:
\begin{equation}\label{ampl}
\chi_\alpha (n)=< \hspace*{-0.3mm}0 \hspace*{-0.5mm}\mid T \hspace*
{-0.5mm} \bigl( \psi (x_1) ... \psi
(x_n)\bigr) \hspace*{-0.7mm}\mid \hspace*{-0.5mm}\alpha \hspace*
{-0.5mm} >\hspace*{0.3mm} ; \
{\bar\chi}_\alpha (n)=< \hspace*{-0.3mm}\alpha \hspace*{-0.5mm}\mid
\hspace*{-0.5mm}T\bigl(\psi (x_1) ... \psi (x_n)\bigr) \hspace*
{-0.7mm}\mid \hspace*{-0.5mm} 0 \hspace*{-0.5mm}> ,
\end{equation}
\begin{equation}\label{prop}
G_{\alpha\beta}(n; m) = \ <0\mid T\biggl( T\bigl(\psi (x_1) ... \psi
(x_n)\bigr) {\hskip.1cm}T\bigl( \psi (x_1^\prime ) ... \psi (x_m^\prime )
\bigr)\biggr) \mid 0 > .
\end{equation}
In order to reduce general forms (\ref{ampl}), (\ref{prop}) to the working
expressions, one needs to perform as well the following steps: \\
\noindent
1. Functional derivation of the $G_{\alpha\beta}(n; m)$. It allows
an expression for the process propagator $G_{\alpha\beta}(n; m)$ through the
cluster propagators $G_i$, $G_f$ and establishes the initial frame for the
cluster decomposition of the process. \\
\noindent
2. Corresponding Gell-Mann, Low limiting procedure
%\cite{gml}
[2].
It assures the QFT consistent contact with a states of the system for
remote times. \\
\noindent
3. Application of the one of the above mentioned procedures (CS, ACS,
or DACS). It results in the segmentation of the scattering process and also
defines the final form of the cluster decomposition.

The exposed steps which are defined for the general propagator imply the
decomposition of the corresponding S-matrix as well as the
corresponding Feyman developement.

We will focus our attention on the third procedure which equips
the system with the corresponding form of the cluster
separability: unspecified, adiabatic or disturbative adiabatic.

{\bf Cluster Separability} is the characteristic that a
system of particles when broken into spatially remote subsystems should be
such  that the dynamics of each subsystem are independent of each
other
%\cite{foldy}-%\cite{sok1}
[3]-[4]. (This property is also denoted as
Separability of the Interaction or Cluster Property.)

The outcome of the CS is a flexible frame for the representation of the
systems dynamics.  In addition it does not introduces any demands for the
"contact processes". A freedom for the interpretation of that part of a
scattering formalism still remains.

The definite demands for the segmentation of the scattering process
are established with the implementation of ACS.

{\bf Adiabatic Cluster Separability}: "No error is introduced into the
treatment of the physically realizable scattering process by a formulation
of the theory in which the interactions among the particles of interest are
'turned off' at remote times-provided, of course,  that  the  turning-off
procedure is sufficiently gradual that  it  does  not,  of  itself,  create
disturbances"
%\cite{kz}
[5]. (The standard name is Adiabatic Hypothesis.)

The configurational content of ACS is a regionalization of the
process in a space-time. The initial (or incoming) region of propagation
${\cal R}_i$, the interaction region ${\cal R}_k$ and the
final (or outgoing) region of propagation  ${\cal R}_f$ are distinguished.
The consequent factorisation of the process representatives (propagator and
S-matrix) is expressed in relations:
\begin{equation}\label{aprop}
G_{\alpha\beta} = G_\beta R_{\alpha\beta}G_\alpha =\prod\limits_{f=1}^lG_fR_
{\alpha\beta}\prod\limits_{i=1}^kG_i ;
\end{equation}
\begin{equation}\label{as-matr}
S_{\beta\alpha}={\bar\chi}_\beta^{out}R_{\alpha\beta}\chi_\alpha^{in}
=\prod\limits_{f=1}^l{\bar\chi}_{\beta
f}R_{\alpha\beta}\prod\limits_{i=1}^k\chi_{\alpha i} .
\end{equation}
In Eqs. (\ref{aprop})-(\ref{as-matr}) $G_\alpha$, $G_\beta$ are channel
propagators, $\chi_\alpha^{in}$, ${\bar\chi}_\beta^{out}$
are the remote times amplitudes,
and $R_{\alpha\beta}$ is a truncated propagator.
$G_i$, $G_f$, ($\chi_{\alpha i}$, ${\bar\chi}_{\beta f}$) are propagators,
(amplitudes) of the scattering process constituents.
A segmentation of the scattering process is expressed on the left-hand side
of the above expressions.
A cluster decomposition is seen on their right-hand side.

A strict application of the ACS implies the sharp distinction of
three propagation regions. Since the propagation is adiabatic, for the whole
regions ${\cal R}_i$ and ${\cal R}_f$ the observable characteristics of the
system are the same as they were or will be at the remote times $\mp\infty$.

We argue that such a feature introduces serious limitations for the
consistent representation of the scattering processes involving
bound states. Particularly critical would be the description
of the intermediate objects: transition operators or segments of the
convolutional expresions  for the form factors, for instance.
Additional problems are related to the interpretation of the contact
processes which are reduced to a cut-off.

In this way one can recognise a limitations for the incorporation of the
ACS into the formalisms which include deconfinement and hadronization.
In particular, it concerns the formalisms which contain on the basic level
different or hybrid levels of compositeness.

We extract three additional groups of motivs for the reformulation
of ACS with a less rigid principle.

I) Motivs of the conceptual nature. The intermediate phases have been
already tacitly introduced (by using, for example, off
mass shell contributions or even intermediate clusterisations of the
channels) without explication whether it is consistent or not with the
general procedure.

II) In applications of the standard formulation one often finds
the incompatibility of the interaction effects originating from scattering
amplitudes and from truncated propagators.

III) The idea was to find consistent and in the same time flexible
formulation which could be applied to many concrete scattering situations.

The rigidity of the ACS and its limitations in encompassing the intermediate
segments led us to the conclusion that one should formulate a
different principle for the segmentation of the scattering processes.

We retained the supposition that interaction between constituents of
scattering processes could be switched off at the remote times, or
precisely for the initial and final space-time regions of propagation.
But in addition we included a new one by which the mechanism of
inclusion/exclusion is such that it creates the disturbances for
initial/final channels in the vicinity of the interaction region.
We supposed also that disturbation effects could be
represented by segments obtained in a factorization of the scattering
picture i.e. that the transition from initial/final regions of
propagation to the region of "pure" interaction, generally, is not direct
but it goes over the intermediate disturbed phases.

{\bf Disturbative Adiabatic Cluster Separability}: In the framework
of the Quantum Field Theory, propagation process can be, generally,
represented by a few-step mechanism. Finite intermediate regions of
disturbance which can be represented by a superposition of the physically
realizable configurations for the corresponding channels are compatibly
connected with the central, interaction region. Interaction among the
constituents of the scattering  process  is  'turned-off'  for  the
initial and final propagation regions, so that initial and final
configurations are corresponding to them
%\cite{olymp}
[1]. (The other name is Disturbative Adiabatic Hypothesis.)

By incorporation of the DACS in the scattering theory space-time content
becomes correspondingly reacher. Two additional domains of scattering
process appear. Disturbed channels of the initial and final configurations
correspond to the initial and final disturbation regions ${\cal
R}_{id}$ i.e. ${\cal R}_{fd}$. So one can visualize the sequence of the
five regions: ${\cal R}_i, {\cal R}_{id}, {\cal R}_k, {\cal R}_{fd}$ and
${\cal R}_f$ separated by the four space-like surfaces. DACS causes the
corresponding factorization of the representatives of the scattering theory.

In our formulation the propagator for the process $\alpha\to\beta$ is written
\begin{equation}\label{daprop}
G_{\alpha\beta}=G_\beta{\bar
K}_\beta^dR_{\alpha\beta}K_\alpha^dG_\alpha=
G_\beta M_{\alpha\beta}G_\alpha .
\end{equation}
\noindent
Here ${\bar K}_\beta^d$ and $K_\alpha^d$ describe the effect of
disturbation introduced. On the right hand side of Eq. (\ref{daprop}) the
expanded interaction term, $M_{\alpha\beta}$,
$M_{\alpha\beta}={\bar K}_\beta^dR_{\alpha\beta}K_\alpha^d$, is defined.

The $S$-matrix gets the form
\begin{equation}\label{das-matr}
S_{\beta\alpha}={\bar\chi}_\beta^{out}{\bar
K}_\beta^dR_{\alpha\beta}K_\alpha^d\chi_\alpha^{in} =
{\bar\chi}_\beta^{out}M_{\alpha\beta}\chi_\alpha^{in} .
\end{equation}

\noindent
The new objects, "disturbation amplitudes" read:
$\chi_\alpha^d\equiv K_\alpha^d\chi_\alpha^{in} ; \
{\bar\chi}_\beta^d\equiv{\bar\chi}_\beta^{out}{\bar K}_\beta^d$.

We derive mathematical basis for the disturbative adibatic (DA)
formalism on the level of the QM S-matrix theory by using the method of
the wave (or M{\o}ller) operators.

The contact with experimental situation is model dependent. The
interaction is given with the form of the expanded interaction term
$M_{\alpha\beta}$. In addition it is necessary to define the disturbation
effects. The remote time amplitudes
$\chi_\alpha^{in}$ and ${\bar\chi}_\beta^{out}$ are determined
phenomenologically.

The general scheme of the DACS could be projected on the various
scattering formalisms. The disturbation input is determined with appropriate
superposition of the physically realizable configurations for the
corresponding channels. The concrete examples of possible applications are
the off mass shell effects in nuclear reactions, the formation of the new
particles states, the transition amplitudes at the quark level and the
appropriate intervention in the Fock expansion.
The interaction content could be described by the corresponding Lagrangian.

The general formalism contains also two procedures which are
independent and mutually compatible.

One of them consist in trying to find "natural" representation of
disturbation effects in the form: \quad
$K^d=G^dV ;\quad {\bar K}^d={\bar V}G^d$.

Propagators $G^d$ are related to the finite domains. They have the same
physical meaning as propagational segments of a truncated propagator
$R_{\alpha\beta}$ which appear in the course of the functional
derivation. $V$, ${\bar V}$ are the vertex type operators defined by the
algoritm. Expanded interaction term $M_{\alpha\beta}$ becomes:
\begin{equation}\label{daintconv}
M_{\alpha\beta}={\bar V}G^dR_{\alpha\beta}G^dV .
\end{equation}
\noindent
One can recognize that the form of the expanded interaction term
(\ref{daintconv}) corresponds to the convolution structure of the
form factor. As an example of this structure one can use the nucleon form
factor described in elastic electron-nucleon scattering within the
QCD hard scattering scheme.

The second procedure consists in the series expansion of $M_{\alpha\beta}$
and reads
\begin{equation}\label{expansion}
M_{\alpha\beta}=\bigl( {\bar K}_\beta^{d(0)}+{\bar
K}_\beta^{d(1)}+...\bigr)\bigl(
R_{\alpha\beta}^{(0)}+R_{\alpha\beta}^{(1)}+...\bigr)\bigl(
K_\alpha^{d(0)}+K_\alpha^{d(1)}+...\bigr) .
\end{equation}
This expansion is not perturbative series but it reflects the content of
Bethe-Salpeter equation with the precise meaning of spatio-temporal
composition of interaction effects.

One needs an additional analysis of terms in (\ref{expansion}) in order
to achieve a compatibility of the interaction contributions independently
of their origin.

The CS, ACS and DACS could be considered as three different
formulations of the general Principle of the segmentation of
scattering process. This principle one could include among the first
principles of the Quantum Field Theory.

The configurational features of this analysis are the part of more general
approach which also includes the studies of covariant propagation and its
correlation to the forms of dynamics. Some segments of that study has been
already reffered
%\cite{fb11}
[6],
%\cite{olymp}
[1].

%% List of equations:
% ampl
% prop
% aprop
% as-matr
% daprop
% das-matr
% daintconv
% expansion

\SaveFinalPage
\end{document}